\documentstyle[12pt,psfig,draft,lscape,array]{nature-ejb}

\def\simlt{\mathrel{\hbox{\rlap{\hbox{\lower4pt\hbox{$\sim$}}}\hbox{$<$}}}}
\def\simgt{\mathrel{\hbox{\rlap{\hbox{\lower4pt\hbox{$\sim$}}}\hbox{$>$}}}}

\def\ale{\mathrel{\hbox{\rlap{\hbox{\lower4pt\hbox{$\sim$}}}\hbox{$<$}}}}
\def\age{\mathrel{\hbox{\rlap{\hbox{\lower4pt\hbox{$\sim$}}}\hbox{$>$}}}}

\def\nodata{---}

\def\ra#1#2#3{#1$^{\rm h}$#2$^{\rm m}$#3$^{\rm s}$}
\def\dec#1#2#3{$#1^\circ#2'#3''$}

\def\grb{GRB\,030329}

\def\spose#1{\hbox to 0pt{#1\hss}}
\newcommand\lsim{\mathrel{\spose{\lower 3pt\hbox{$\mathchar"218$}}
     \raise 2.0pt\hbox{$\mathchar"13C$}}}
\newcommand\gsim{\mathrel{\spose{\lower 3pt\hbox{$\mathchar"218$}}
     \raise 2.0pt\hbox{$\mathchar"13E$}}}

\begin{document}

\title{\Large \bf A Common Origin for Cosmic Explosions Inferred from
Fireball Calorimetry}

\author{
   E.~Berger\affiliation[1]
     {Caltech Optical Observatories 105-24, California Institute of 
     Technology, Pasadena, CA\,91125, USA},
   S.~R.~Kulkarni\affiliationmark[1],
   G.~Pooley\affiliation[2]
     {Mullard Radio Astronomy Observatory, Cavendish Lab.,
     Madingley Road, Cambridge CB3 0HE, UK},
   D.~A.~Frail\affiliation[3]
     {National Radio Astronomy Observatory, P.O. Box 0, Socorro, New
     Mexico 87801, USA},
   V.~McIntyre\affiliation[4]
     {Australia Telescope National Facility, CSIRO, P.O. Box 76, 
      Epping, NSW 1710, Australia},
   R.~M.~Wark\affiliation[5]
     {Australia Telescope National Facility, CSIRO, Locked Bag 194, Narrabri 
      NSW 2390, Australia},
   R.~Sari \affiliation[6]
     {Theoretical Astrophysics 130-33, California Institute of
     Technology, Pasadena, CA\,91125, USA}
   A.~M.~Soderberg\affiliationmark[1],
   D.~W.~Fox\affiliationmark[1],
   S.~Yost\affiliation[7]
     {Space Radiation Laboratory 220-47, California Institute of 
     Technology, Pasadena, CA\,91125, USA}, 
   P.~A.~Price\affiliation[8]
     {RSAA, ANU, Mt.\ Stromlo Observatory, via Cotter Rd, Weston
     Creek, ACT, 2611, Australia}
}

\date{\today}{}
\headertitle{Calorimetry of a Nearby GRB}
\mainauthor{Berger et al.}

\summary{Past studies \cite{fks+01,bkf03,bfk03} suggest that 
long-duration $\gamma$-ray bursts (GRBs) have a standard energy of
$E_\gamma\sim 10^{51}$ erg in ultra-relativistic ejecta when corrected
for asymmetry (``jets'').  However, recently \cite{bkf03,bfk03} a
group of sub-energetic bursts, including the peculiar GRB\,980425
associated \cite{gal+98} with SN\,1998bw ($E_\gamma\approx 10^{48}$
erg), has been identified.  Here we report radio observations of
\grb, the nearest burst to date, which allow us to undertake
calorimetry of the explosion.  Our observations require a
two-component explosion: a narrow ($5^\circ$) ultra-relativistic
component responsible for the $\gamma$-rays and early afterglow, and a
wide, mildly relativistic component responsible for the radio and
optical afterglow beyond 1.5 days.  While the $\gamma$-rays are
energetically minor, the total energy release, dominated by the wide
component, is similar \cite{fks+01,bkf03,bfk03,pk02} to that of other
GRBs.  Given the firm link \cite{smg+03,hjo+03} of GRB\,030329 with
SN\,2003dh our result suggests a common origin for cosmic explosions
in which, for reasons not understood, the energy in the highest
velocity ejecta is highly variable.}

\maketitle

%%%%%%%%%%%%%%%%%%%%%%%%%%%%%%%%%%%%%%%%%%%%%%%%%%%%%%%%%%%%%%%%%%%

\begin{figure}
\centerline{\psfig{file=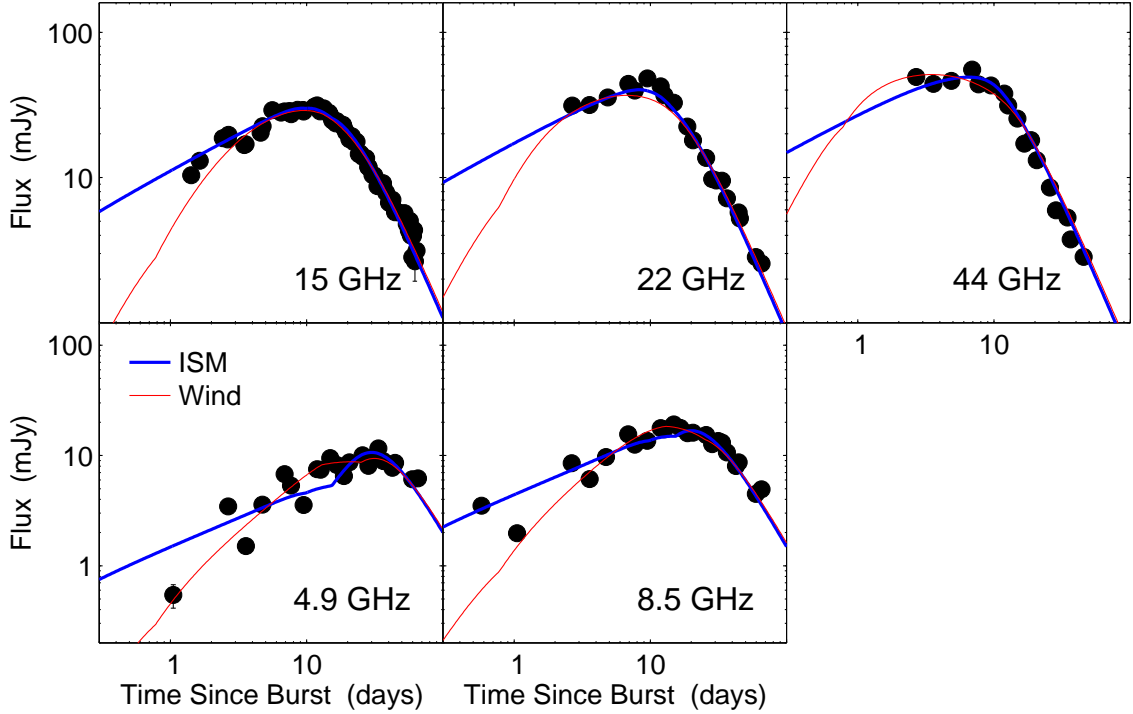,width=6in,angle=0}}
\caption[]{\small Radio lightcurves of the afterglow of \grb{}.  
All measurements
include $1\sigma$ error bars which in most cases are smaller than the
symbols.  The data are summarized in Tables~\ref{tab:vla} and
\ref{tab:ryle}.  The solid lines are models of synchrotron emission
from collimated relativistic ejecta expanding into uniform
\cite{sph99} (thick) and wind \cite{cl00} $\rho\propto r^{-2}$ (thin)
circumburst media; these models include the millimetre and
sub-millimetre data \cite{sfw+03}.  We find $\chi^2_r=31.3$ and $39.8$
(164 degrees of freedom) for the uniform density and wind models,
respectively; these include a $2\%$ systematic error added in
quadrature to each measurement.  The large values of $\chi^2_r$ are
dominated by interstellar scintillation (ISS) at $\nu\simlt 15$ GHz
and mild deviations from the expected smooth behavior at the high
frequencies.  Comparing the data and models, we find rms flux
modulations of 0.25 at 4.9 GHz, 0.15 at 8.5 GHz, and 0.08 at 15 GHz,
as well as a drop by a factor of three in the level of modulation from
$\sim 3$ to 40 days.  These properties are expected in weak ISS as the
fireball expands on the sky.  The inferred source size of about 20
$\mu$as (i.e.~$\sim 2\times 10^{17}$ cm) at $t\sim 15$ days is in
close agreement with theoretical expectations \cite{gfs+03}.  In the
uniform density model the jet break occurs at $t\approx 10$ d
corresponding to an opening angle, $\theta_j\approx 0.3$ ($17^\circ$).
From the derived synchrotron parameters (at $t=t_j$): $\nu_a\approx
19$ GHz, $\nu_m\approx 43$ GHz, $F_{\nu,m}\approx 96$ mJy we find an
isotropic kinetic energy, $E_{K,{\rm iso}}\approx 5.6\times10^{51}
\nu_{c,13}^{1/4}$ erg, a circumburst density $n=1.8\nu_{c,13}^{3/4}$
cm$^{-3}$, and the fractions of energy in the relativistic electrons
and magnetic field of $0.16\nu_{c,13}^{1/4}$ and
$0.10\nu_{c,13}^{-5/4}$, respectively; here $\nu_c=10^{13}\nu_{c,13}$
is the synchrotron cooling frequency, and a constraint on Inverse
Compton cooling as advocated by Sari \& Esin \cite{se01} indicates
$\nu_{c,13}\simlt 1$.  The beaming-corrected kinetic energy is
$E_K\approx 2.5\times 10^{50} \nu_{c,13}^{1/4}$ erg, typical of other
well-studied long-duration GRBs \cite{pk02}.  The parameters derived
from the wind model are consistent with those from the uniform density
model to within $10\%$.}
\label{fig:lcs}
\end{figure}

We initiated observations of the nearby \grb{} ($z=0.1685$) in the
centimetre band approximately 13.8 hours after the burst.  The log of
the observations and the resulting lightcurves are displayed in
Tables~\ref{tab:vla} and \ref{tab:ryle} and Figure~\ref{fig:lcs}.  The
afterglow was also observed extensively in the millimetre (100 GHz) and
sub-millimetre (250 GHz) bands \cite{sfw+03}.  While this is the brightest 
radio afterglow detected to date, the low redshift results in a peak
luminosity, $L_{\nu,p} (8.5\,{\rm GHz}) \approx 1.8\times 10^{31}$ erg
s$^{-1}$ Hz$^{-1}$, typical \cite{fkb+03} of other long-duration GRBs.

The observed rapid decline, $F_\nu\propto t^{-1.9}$ at $t\simgt 10$ d
and the decrease in peak flux at $\nu\simlt 22.5$ GHz
(Figure~\ref{fig:lcs}) are the hallmarks of a collimated explosion.
In this framework \cite{sph99}, the sharp decline (or ``jet break'')
occurs at the time, $t_j$, when $\Gamma(t_j)\sim\theta_j^{-1}$ due to
relativistic abberation (``beaming'') and rapid side-ways expansion;
here $\Gamma$ is the bulk Lorentz factor and $\theta_j$ is the opening
angle of the jet.  We model the afterglow emission
(cf.~ref.~\pcite{pk02},\pcite{bsf+00}) from 4.9 to 250 GHz assuming a
uniform \cite{sph99} as well as a ``wind'' \cite{cl00} (particle
density profile, $\rho\propto r^{-2}$, where $r$ is the distance from
the source) circumburst medium.  Neither model is strongly preferred,
but $t_{j,\rm rad}\approx 9.8\,$d is required (Figure~\ref{fig:lcs}).

Using the inferred particle density of $n\approx 1.8$ cm$^{-3}$ and
assuming a $\gamma$-ray efficiency, $\epsilon_\gamma=0.2$ (see
ref.~\pcite{bfk03}) we infer $\theta_{j,{\rm rad}}\sim 0.3$ rad, or
17$^\circ$.  The kinetic energy in the explosion corrected for
collimation is $E_K=f_bE_{K,{\rm iso}}\approx 2.5\times 10^{50}$ erg,
where $f_b=[1-\cos(\theta_j)]$ is the beaming fraction and $E_{K,{\rm
iso}}$ is the isotropic equivalent kinetic energy.  This value is
comparable to that inferred from modeling of other afterglows
\cite{pk02}.

\begin{figure}
\centerline{\psfig{file=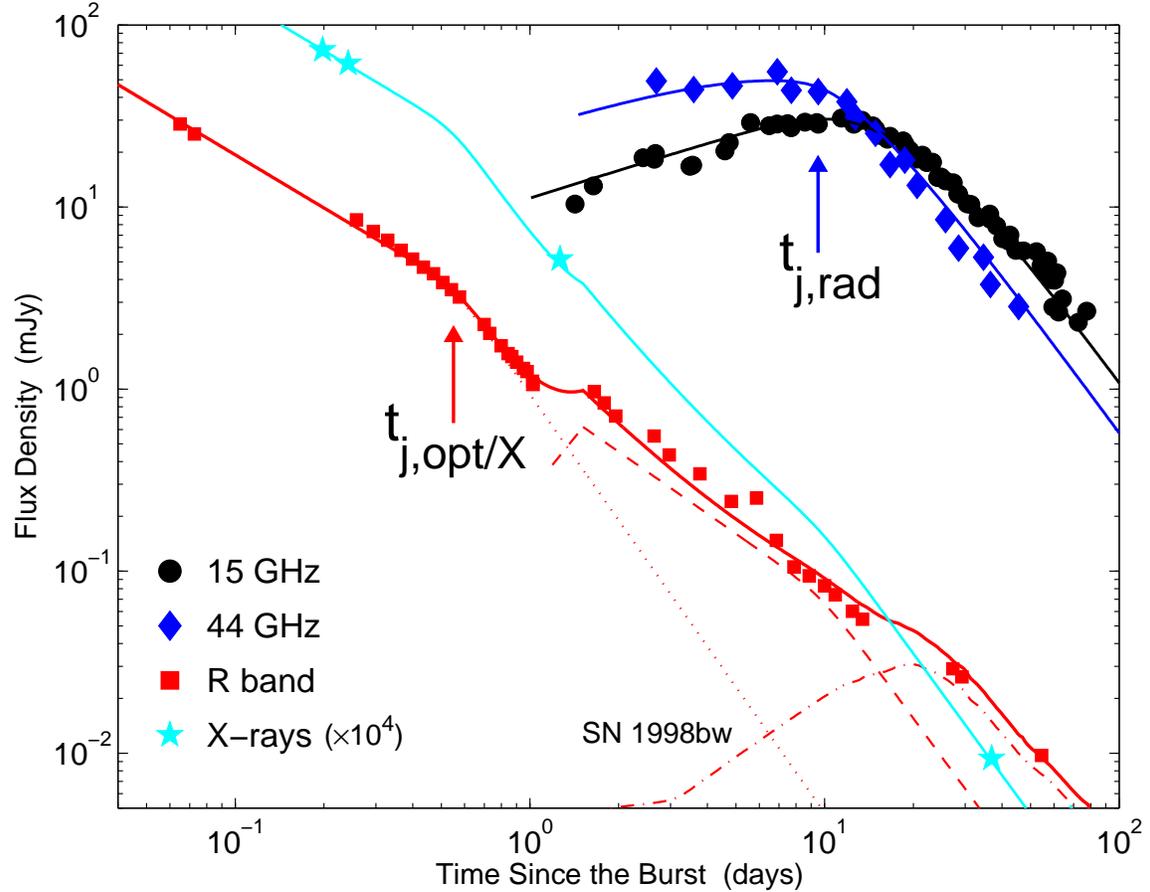,width=6in,angle=0}}
\caption[]{\small Radio to X-ray lightcurves of the afterglow of
\grb{}.  The optical data, from Price et al.~\cite{pfk+03} and the GRB
Coordinates Network \cite{hcz+03,iak+03b,tcm+03}, have been corrected
for Galactic extinction, $A_R=0.067$ mag; we note that the latter are
preliminary data.  The dotted line is the model proposed by Price et
al. \cite{pfk+03} for the early optical emission, with $t_{j,\rm
opt}\approx 0.55$ d.  The dashed line is an extapolation of the
uniform density model presented in Figure~\ref{fig:lcs} to the optical
$R$-band with $\nu_{c,13}=2$; this value is somewhat larger than the
rough limit discussed in Figure~\ref{fig:lcs} but may be consistent
with the uncertainty in the model parameters.  The model in the X-ray
band is based on the measured \cite{tmg+03} optical to X-ray spectral
slope and an extapolation of the uniform density model presented in
Figure~\ref{fig:lcs}.  The sharp increase in the optical flux at
$t\simlt 1.5$ d is due to the deceleration of the slower second jet
component.  Finally, the dot-dashed line is the optical emission from
SN\,1998bw at the redshift of \grb{}, $z=0.1685$, used as a proxy for
SN\,2003dh \cite{smg+03}.  The solid line in the $R$-band is a
combination of the SN and the two jet components, whereas in the radio
and millimetre bands it is the uniform density model presented in
Figure~\ref{fig:lcs}.  In the X-ray band the model is dominated by the
narrow jet component.  While this two-component jet model provides a
reasonable fit to the data, there are still some discrepancies which
can be resolved by accurate photometry and a more careful subtraction
of SN\,2003dh.  The latter will also allow a more precise
determination of the putative second jet break in the optical band at
$t_{j,\rm rad}$.}
\label{fig:ro}
\end{figure}

In contrast to the above discussion, Price et al. \cite{pfk+03} note a
sharp break in the optical afterglow at $t=0.55$ d
(Figure~\ref{fig:ro}).  The X-ray flux \cite{tmg+03} tracks the
optical afterglow for the first day, with a break consistent with that
seen in the optical.  Thus the break at 0.55 d is not due to a change
in the ambient density since for typical parameters \cite{kum00,fw01} 
the X-ray emission is not sensitive to density.  However, unlike the
optical emission the X-ray flux at later times continues to decrease
monotonically.  Thus we conclude that there are two emitting
components: one responsible for the early optical and X-ray emission
and the other responsible for the optical emission beyond 1.5 days.

The first component, given the characteristic $t^{-2}$ decay for both
the X-ray and optical emission, is reasonably modeled by a jet.  For
the parameters used above ($n$, $\epsilon_\gamma$) the opening angle
is 0.09 rad or $5^\circ$.

The resurgence in the optical emission at 1.5 d requires a second
component.  An increase in the ambient density cannot explain this
resurgence since the predicted decrease in radio luminosity, arising
from the increase in synchrotron self-absorption, is not observed
(Figure~\ref{fig:lcs}).  An increase in the energy of the first
component, for example by successive shells with lower Lorentz factors
as advocated by Granot et al. \cite{gnp03}, is ruled out by the lack
\cite{sfw+03} of strong radio or millimetric emission expected
\cite{sm00} from reverse shocks.

Thus, by a process of elimination, we are led to a two-component
explosion model in which the first component (a narrow jet, $5^\circ$)
with initially larger $\Gamma$ is responsible for the $\gamma$-ray
burst and the early optical and X-ray afterglow including the break at
0.55 d, while the second component (a wider jet, $17^\circ$) powers
the radio afterglow and late optical emission (Figure~\ref{fig:ro}).
The break due to the second component is readily seen in the radio
afterglow, but is masked by SN\,2003dh in the optical bands, thus
requiring careful subtraction (Figure~\ref{fig:ro}).  Such a
two-component jet finds a natural explanation in the collapsar model
\cite{mwh01}.

The beaming-corrected $\gamma$-ray energy, emitted by the narrow jet,
is only $E_\gamma\approx 5\times 10^{49}$ erg, significantly lower
than the strong clustering \cite{bfk03} around $1.3\times 10^{51}$ erg
seen in most bursts.  Similarly, the beaming-corrected X-ray
luminosity \cite{tmg+03} at $t=10$ hours, a proxy for the kinetic
energy of the afterglow on that timescale, is $L_{X,10}\approx 3\times
10^{43}$ erg s$^{-1}$, a factor of ten below the tightly clustered
values \cite{bkf03} for most other bursts.  However, the second
component, which is mildly relativistic (as determined by the lower
energy peak of its spectrum), carries the bulk of the energy, as
indicated by our modeling of the radio emission.  We note that our
model, with the energy in the lower Lorentz factor component
dominating over the narrow ultra-relativistic component, is not
consistent with ``universal standard jet'' model \cite{rlr02}.

\begin{figure}
\centerline{\psfig{file=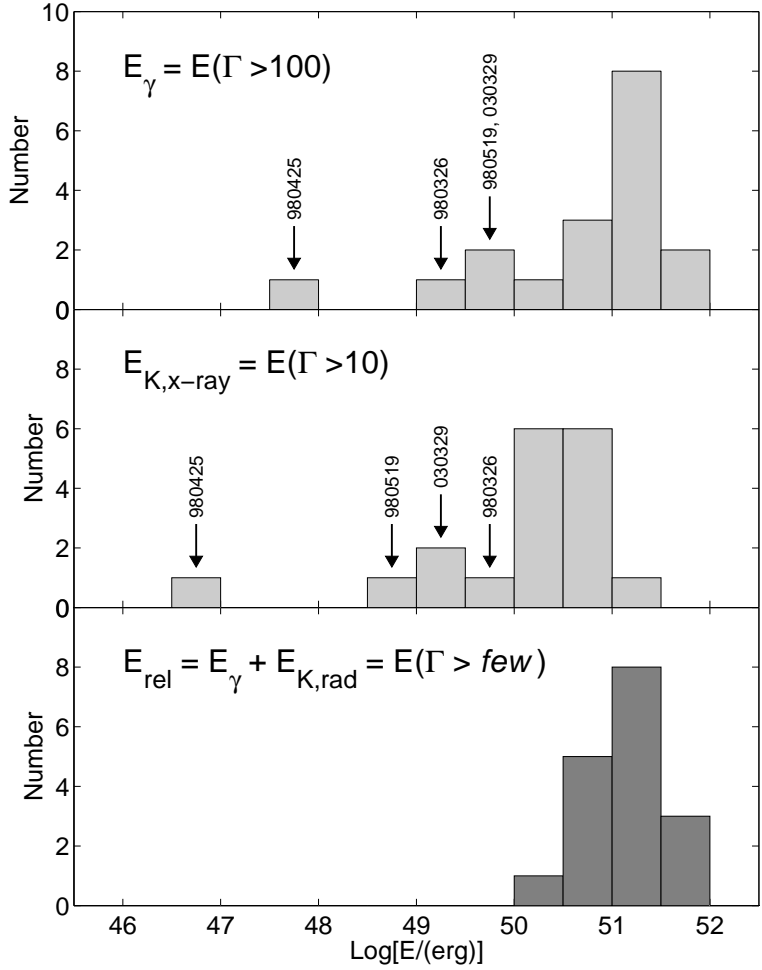,width=4in,angle=0}}
\caption[]{\small Histograms of various energies measured for GRBs.
We plot the beaming-corrected $\gamma$-ray
energy \cite{bfk03}, $E_\gamma$, the kinetic energy inferred from
X-rays at $t=10$ hr \cite{bkf03}, $E_{K,X}$, and total
relativistic energy, $E_{\rm rel}=E_\gamma+E_{K}$, where $E_K$ is the
beaming-corrected kinetic energy inferred \cite{lc99,pk02} from the
broad-band afterglow.  The energy in X-rays, $E_{K,X}=L_Xt/
\epsilon_e(\alpha_X-1)$, with $t=10$ hr, $\epsilon_e=0.1$, and
$\alpha_X=1.3$ is the median decay rate in the X-ray band.  For
GRB\,980519 we find that the evolution of the radio emission requires
a much wider jet, $\theta_j\sim 0.3$, than what is inferred from the
optical, $\theta_j\sim 0.05$; here we assume $z=1$.  We therefore
infer $E_K\sim 2\times 10^{50}$ erg from the radio data compared to
$E_\gamma\approx 4\times 10^{49}$ erg.  The $\gamma$-ray energy of
GRB\,980425 is an upper limit since the degree of collimation is not
known.  For the kinetic energy we use the value derived by Li \&
Chevalier \cite{lc99} based on the radio evolution of SN\,1998bw.
There is a significantly wider dispersion in $E_\gamma$ and $E_{K,X}$
as compared to the total explosive yield.  This indicates that engines
in cosmic explosions produce approximately the same quantity of energy
thus pointing to a common origin, but the quality of these engines, as
indicated by ultra-relativistic output, varies widely.}
\label{fig:energy}
\end{figure}

The afterglow calorimetry presented here has important ramifications
for our understanding of GRB engines.  Recently, we have come to
recognize a sub-class of cosmological GRBs marked by rapidly fading
afterglows at early time (i.e.~similar to GRB\,030329).  These events
are sub-energetic \cite{bkf03,bfk03} in $E_\gamma$ and early X-ray
afterglow luminosity.  However, as demonstrated by our calorimetry of
GRB\,030329, such bursts may have total explosive yields similar to
other GRBs (Figure~\ref{fig:energy}).

This leads to the following conclusions. First, radio calorimetry,
which is sensitive to all ejecta with $\Gamma\simgt few$, shows that
the explosive yield of the nearest ``classical'' event,
\grb, is dominated by mildly relativistic ejecta. Ultra-relativistic
ejecta which produced the $\gamma$-ray emission is energetically
unimportant.  Second, the total energy yield of \grb{} is similar to
those estimated for other bursts.  Along these lines, the enigmatic
GRB\,980425 associated \cite{gal+98} with the nearby supernova
SN\,1998bw also has negligible $\gamma$-ray emission, $E_{\gamma,{\rm
iso}}\approx 8\times 10^{47}$ erg; however, radio calorimetry
\cite{lc99} shows that even this extreme event had a similar explosive
energy yield (Figure~\ref{fig:energy}).  The newly recognized class of
cosmic explosions, the X-ray Flashes \cite{hzk03}, exhibit little or no
$\gamma$-ray emission but appear to have comparable X-ray and radio
afterglows to those of GRBs.  Thus, the commonality of the total
energy yield indicates a common origin, but apparantly the
ultra-relativistic output is highly variable.  Unraveling what
physical parameter is responsible for the variation in the ``purity''
(ultra-relativistic output) of the engine appears to be the next
frontier in the field of cosmic explosions.

\begin{acknowledge}
GRB research at Caltech is supported in part by funds from NSF and
NASA.  We are, as always, indebted to Scott Barthelmy and the GCN.
The VLA is operated by the National Radio Astronomy Observatory, a
facility of the National Science Foundation operated under cooperative
agreement by Associated Universities, Inc. The Australia Telescope is
funded by the Commonwealth of Australia for operations as a National
Facility managed by CSIRO.  The Ryle Telescope is supported by PPARC. 

\end{acknowledge}

\clearpage
\begin{table}
\begin{center}
\setlength{\extrarowheight}{-0.075in}
\begin{tabular}{>{\scriptsize}l >{\scriptsize}c >{\scriptsize}c >{\scriptsize}c >{\scriptsize}c >{\scriptsize}c >{\scriptsize}c >{\scriptsize}c}
\hline
\hline
Epoch & $\Delta t$ & $F_{1.43}$ & $F_{4.86}$ & $F_{8.46}$ &
$F_{15.0}$ & $F_{22.5}$ & $F_{43.3}$ \\
(UT) & (days) & (mJy) & (mJy) & (mJy) & (mJy) & (mJy) & (mJy) \\
\hline 
Mar 30.06 & 0.58 & \nodata & \nodata & $3.50\pm 0.06$ & \nodata &
\nodata & \nodata \\
Mar 30.53 & 1.05 & \nodata & $0.54\pm 0.13$ & $1.98\pm 0.17$ & \nodata
& \nodata & \nodata \\
Apr 1.13 & 2.65 & $<0.21$ & $3.45\pm 0.05$ & $ 8.50\pm 0.05$ &
$19.68\pm 0.14$ & $30.40\pm 0.06$ & $46.63\pm 0.18$ \\
Apr 2.05 & 3.57 & $<0.30$ & $ 1.51\pm 0.05$ & $6.11\pm 0.04$ &
$16.98\pm 0.19$ & $31.59\pm 0.14$ & $44.17\pm 0.35$  \\
Apr 3.21 & 4.76 & $<0.36$ & $ 3.58\pm 0.04$ & $ 9.68\pm 0.03$ &
$22.59\pm 0.12$ & $35.57\pm 0.09$ & $46.32\pm 0.23$  \\
Apr 5.37 & 6.89 & $<0.40$ & $ 6.77\pm  0.08$ & $15.56\pm 0.06$
& $28.58\pm 0.20$ & $44.09\pm 0.15$ & $55.33\pm 0.43$ \\
Apr 6.16 & 7.68 & $<0.25$ & $ 5.34\pm 0.10$ & $12.55\pm 0.21$ &
$27.26\pm 0.21$ & $39.68\pm 0.20$ & $43.81\pm 1.00$ \\
Apr 7.97 & 9.49 & $<0.68$ & $ 3.55\pm 0.11$ & $13.58\pm 0.09$ &
$28.50\pm 0.23$ & $48.16\pm 0.23$ & $43.06\pm 1.33$  \\
Apr 10.38 & 11.90 & $<0.58$ & $ 7.51\pm 0.08$ & $17.70\pm 0.05$ & 
$31.40\pm 0.25$ & $42.50\pm 0.14$ & $37.86\pm 0.46$  \\
Apr 11.17 & 12.69 & \nodata & $ 7.42\pm  0.09$ & $17.28\pm  0.10$ & 
$29.60\pm 0.29$ & $36.84\pm 0.16$ & $31.26\pm 0.51$ \\
Apr 13.35 & 14.87 & \nodata & $ 9.49\pm 0.13$ &  $19.15\pm 0.08$ &
$26.78\pm 0.33$ & $32.69\pm 0.13$ & $25.44\pm 0.51$ \\
Apr 15.14 & 16.66 & \nodata & $ 8.21\pm 0.08$ &  $17.77\pm 0.10$ &
$24.50\pm 0.31$ & \nodata & $17.10\pm 0.71$ \\
Apr 17.20 & 18.72 & $<0.63$ & $ 6.50\pm 0.11$ & $15.92\pm 0.07$ &
$22.02\pm 0.25$ & $22.41\pm 0.08$ & $18.07\pm 0.28$ \\
Apr 19.06 & 20.58 & \nodata & $ 8.66\pm 0.10$ & $16.08\pm 0.06$ &
$18.35\pm 0.24$ & $18.03\pm 0.11$ & $13.15\pm 0.29$  \\
Apr 24.18 & 25.70 & \nodata & $10.04\pm 0.08$ & $15.34\pm 0.06$ &
$13.93\pm 0.26$ & $13.63\pm 0.13$ & $ 8.54\pm 0.48$  \\
Apr 26.92 & 28.44 & $<0.58$ & $ 8.05\pm 0.08$ & $12.67\pm 0.09$ &
$11.82\pm 0.26$ & $ 9.75\pm 0.23$ & $ 5.95\pm 0.62$  \\
Apr 28.96 & 30.48 & \nodata & \nodata & \nodata & $10.40\pm 0.33$ &
$9.53\pm 0.21$ & \nodata \\
Apr 29.99 & 31.51 & \nodata & $ 9.80\pm 0.09$ & $13.55\pm 0.07$ &
\nodata & \nodata & \nodata \\
May 2.06 & 33.58 & \nodata & $11.62\pm 0.08$ & $13.10\pm 0.06$ &
\nodata & $ 9.52\pm 0.14$ & \nodata \\
May 3.07 & 34.59 & \nodata & \nodata & \nodata & \nodata & \nodata &
$5.30\pm 0.32$ \\
May 5.00 & 36.52 & \nodata & $8.90\pm 0.08$ & $10.64\pm 0.06$ &
$8.58\pm 0.17$ & $7.20\pm 0.09$ & $3.75\pm 0.26$ \\
May 11.03 & 42.55 & \nodata & $7.72\pm 0.13$ & $8.04\pm 0.08$ &
$7.03\pm 0.19$ & \nodata & \nodata \\
May 13.03 & 44.55 & \nodata & $8.57\pm 0.09$ & $8.68\pm 0.08$ &
$5.77\pm 0.22$ & $5.75\pm 0.10$ & \nodata \\
May 14.00 & 45.52 & \nodata & \nodata & \nodata & \nodata & $5.23\pm
0.17$ & $2.84\pm 0.23$ \\
May 28.03 & 59.55 & \nodata & $6.08\pm 0.10$ & $4.48\pm 0.09$ &
$2.82\pm 0.21$ & $2.84\pm 0.20$ & \nodata \\
June 4.01 & 66.53 & $1.94\pm 0.06$ & $6.20\pm 0.08$ & $4.93\pm 0.06$ &
\nodata & $2.56\pm 0.12$ & \nodata
\end{tabular}
\end{center}
\caption[]{\small Radio observations made with the Very Large Array 
(VLA) and the Australia Telescope Compact Array (ATCA).  Observations
commenced on March 30.06 UT with a single 7-hour observation with 
ATCA on Mar.~30.53 UT.  In the initial observation we detected a point 
source at $\alpha$(J2000)=\ra{10}{44}{49.95}, 
$\delta$(J2000)=\dec{21}{31}{17.38}, with an uncertainty of about 0.1
arcsec in each coordinate, consistent with the position of the optical
counterpart.  In all VLA observations we used the standard continuum
mode with $2\times 50$ MHz bands.  At 22.5 and 43.3 GHz we used
referenced pointing scans to correct for the systematic $10-20$ arcsec
pointing errors of the VLA antennas.  We used the extra-galactic
sources 3C\,147 (J0542+498) and 3C\,286 (J1331+305) for flux
calibration, while the phase was monitored using J1111+199 at 1.43 GHz
and J1051+213 at all other frequencies.  The ATCA observations were
performed at 4.80, 6.21, 8.26, and 9.02 GHz with a bandwidth of 64 MHz
in each frequency.  The phase was monitored using J1049+215, while the
flux was calibrated using J1934-638.  The data were reduced and
analyzed using the Astronomical Image Processing System (VLA) and the
Multichannel Image Reconstruction, Image Analysis and Display package
(ATCA).  The flux density and uncertainty were measured from the
resulting maps by fitting a Gaussian model to the afterglow.  In
addition to the rms noise in each measurement we estimate a systematic
uncertainty of about 2\% due to uncertainty in the absolute flux
calibration.}
\label{tab:vla}
\end{table}

\clearpage

\begin{table}
\begin{center}
\setlength{\extrarowheight}{-0.075in}
\begin{tabular}{>{\scriptsize}l >{\scriptsize}c >{\scriptsize}c ||
>{\scriptsize}l >{\scriptsize}c >{\scriptsize}c}
\hline
\hline
\small Epoch & $\Delta t$ & $F_{15.3}$ & Epoch & $\Delta t$ & $F_{15.3}$ \\
(UT) & (days) & (mJy) & (UT) & (days) & (mJy) \\ \hline
Mar 30.91 & 1.43 & $10.38\pm 0.28$  & Apr 21.72 & 23.24 & $17.63\pm 0.29$  \\
Mar 31.12 & 1.64 & $13.05\pm 0.28$  & Apr 22.66 & 24.18 & $14.51\pm 0.49$  \\
Mar 31.91 & 2.43 & $18.66\pm 0.28$  & Apr 23.33 & 24.85 & $14.62\pm 0.49$  \\
Apr 1.12 & 2.64 & $18.29\pm 0.28$   & Apr 25.81 & 27.33 & $13.60\pm 0.65$  \\
Apr 1.98 & 3.50 & $16.75\pm 0.27$   & Apr 26.82 & 28.34 & $11.78\pm 0.52$  \\
Apr 3.07 & 4.59 & $20.36\pm 0.45$   & Apr 29.82 & 31.34 & $10.35\pm 0.49$  \\
Apr 4.09 & 5.61 & $29.13\pm 0.52$   & May 1.63 & 33.15 & $8.73\pm 0.52$   \\
Apr 4.97 & 6.49 & $27.97\pm 0.26$   & May 4.80 & 36.32 & $9.15\pm 0.50$   \\
Apr 5.97 & 7.49 & $28.69\pm 0.26$   & May 6.83 & 38.35 & $7.87\pm 0.50$   \\
Apr 7.06 & 8.58 & $29.29\pm 0.49$   & May 8.73 & 40.25 & $6.70\pm 0.50$   \\
Apr 7.89 & 9.41 & $29.15\pm 0.44$   & May 10.76 & 42.28 & $6.49\pm 0.50$  \\
Apr 9.89 & 11.41 & $30.78\pm 0.51$  & May 15.76 & 47.28 & $5.74\pm 0.50$  \\
Apr 11.05 & 12.57 & $28.52\pm 0.51$ & May 20.70 & 52.22 & $5.69\pm 0.53$ \\
Apr 11.88 & 13.40 & $29.92\pm 0.44$ & May 22.76 & 54.28 & $4.78\pm 0.78$ \\ 
Apr 13.05 & 14.57 & $27.90\pm 0.44$ & May 24.76 & 56.28 & $4.31\pm 0.55$ \\
Apr 13.87 & 15.39 & $24.74\pm 0.44$ & May 25.56 & 57.08 & $5.04\pm 0.84$ \\
Apr 14.82 & 16.34 & $23.60\pm 0.32$ & May 26.75 & 58.27 & $3.99\pm 0.63$ \\
Apr 16.96 & 18.48 & $23.06\pm 0.24$ & May 28.76 & 60.28 & $3.96\pm 0.58$ \\
Apr 17.92 & 19.44 & $20.51\pm 0.24$ & May 29.82 & 61.34 & $4.35\pm 0.50$ \\
Apr 19.95 & 21.47 & $19.27\pm 0.38$ & May 30.76 & 62.28 & $2.65\pm 0.72$ \\
Apr 20.72 & 22.24 & $17.53\pm 0.33$ & June 2.54 & 64.06 & $3.13\pm 0.76$
\end{tabular}
\end{center}
\caption[]{\small Radio observations at 15.3 GHz made with the Ryle 
Telescope at Cambridge (UK).   All observations were made by interleaving 
15 minutes scans of \grb{}
with 2.5 minutes scans of the phase calibrator J1051+2119.  The
absolute flux scale was calibrated using 3C\,48 and 3C\,286.  We used
5 antennas providing 10 baselines in the range 35 -- 140 m.  Since the
position of the source is well known the in-phase component of the
vector sum of the 10 baselines was used as an unbiased estimate of the
flux density.  The typical rms fluctuation on the signal in a 32-s
integration period is approximately 6 mJy.  We also add a systematic
uncertainty of about 2\% due to uncertainty in the absolute flux
calibration.}
\label{tab:ryle}
\end{table}

\end{document}